# Monitoring hemolysis continuously in real time

Tyler Van Buren,[1*] Gilad Arwatz,[2] Alexander J. Smits[1]

[1]Mechanical and Aerospace Engineering Department, Princeton University
Princeton, NJ 08544, USA

[2]Instrumems Inc.
Sunnyvale, CA 94805, USA

[*]To whom correspondence should be addressed; E-mail: tburen@princeton.edu.

**Blood damage (hemolysis) can occur during clinical procedures, e.g. dialysis, due to human error or faulty equipment, and it can cause significant harm to the patient or even death. We propose a simple technique to monitor changes in hemolysis levels accurately, continuously, and in real time. As red blood cells rupture, the overall conductivity of the blood increases. Here, we demonstrate that small changes in porcine blood hemolysis can be detected accurately through a simple resistance measurement.**

## Introduction

Hemolysis (the rupture of red blood cells) can occur in medical procedures where blood is removed from the body (*1, 2*). For example, passing blood through a faulty dialysis machine can potentially risk the life of the patient (*3, 4*), and even drawing blood too quickly through a needle can lead to defective laboratory samples (*5*). Preventing hemolysis is therefore an important design constraint for medical pumps, prosthetic organs, hypodermic needles, and blood extraction procedures (*6–9*).

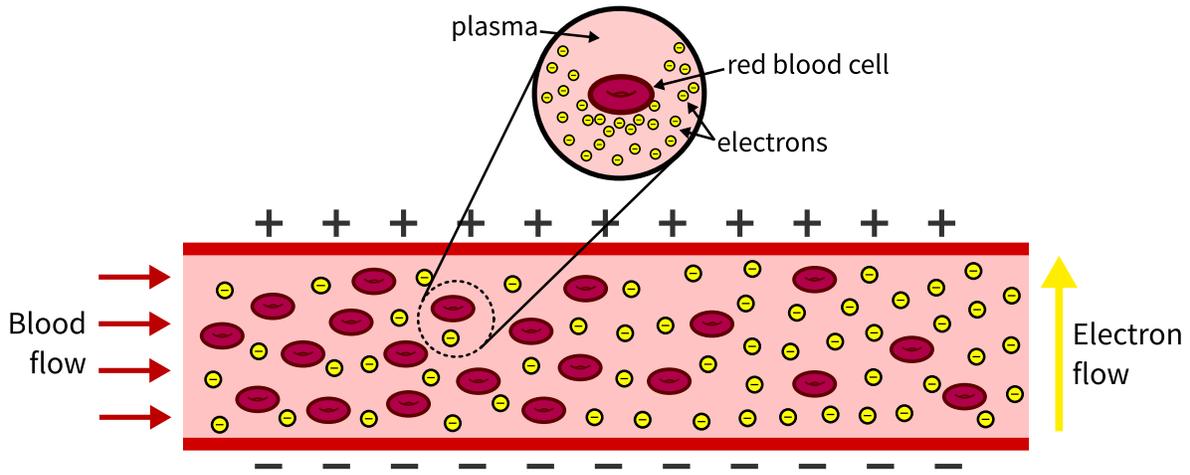

Figure 1: Simplified illustration of blood flow to show how hemolysis leads to a higher blood conductivity. The fluid becomes more conductive left-to-right as red blood cells rupture and release their hemoglobin.

The cytoplasm of red blood cells is rich in hemoglobin, an iron-containing biomolecule that can bind oxygen and is responsible for the red color of the cells. As red blood cells rupture, they release their hemoglobin into the plasma— which is mostly water— changing the plasma from being relatively colorless to having a red tint. The degree of hemolysis can be measured by separating the plasma from the red blood cells and analyzing the amount of cell-free hemoglobin (*10*) using a spectrophotometer, which measures how much light of a given wavelength is absorbed by the sample. Spectrophotometry is considered the most accurate method for measuring hemolysis, but there are a number of other possible methods that do not require extracting red blood cells or using chemical analysis. For example, Tarasev (*11*) suggested a blood hemolysis analyzer that can measure the amount of cell-free hemoglobin by using two or more wavelengths of light and comparing the different levels of light absorption. Karlsson (*12*) and Lee (*13*) both describe devices to identify hemolysis by the naked eye through the change in plasma color. None of these methods, however, allow clinicians to monitor hemolysis during a procedure and provide immediate information on the level of blood damage.

Recently, Zhou et al. (*14*) proposed a complex method that could actively measure hemolysis by combining nanofilters, which actively filter the plasma from the red blood cells, and optofluidic sensors for evanescent absorption detection. The process is similar to spectrophotometry, in that it is analyzing light absorption and relating it to hemolysis level, with the added benefit of eliminating the need for specialty sample preparation.

We propose a much simpler technique that can detect hemolysis continuously in real-time by measuring the electrical resistance of the blood. Blood is naturally conductive, but the outer lipid bilayer of the red blood cell is insulating and so healthy blood cells do not contribute much to the overall conductivity (*15, 16*). As red blood cells rupture, however, they release their hemoglobin and raise the conductivity of the entire fluid (as illustrated in figure 1). Hence, the change in blood conductivity can be related directly to the level of hemolysis. Our proposed technique does not require external sources of light, the separation of the blood cells from the plasma, or specific chemical detection, yet it can determine the progression of hemolysis in real time.

## Results and discussion

The method uses a test cell that consists of a small converging/diverging channel equipped with top and bottom electrodes. The cell was tested by inserting it in a laminar flow loop driven by a peristaltic pump, as shown in figure 2. The conductivity was measured using a high-quality inductance-capacitance-resistance (LCR) meter for continuous sampling, and a conventional conductivity probe for periodic sampling.

The system was first tested using KCl saline solutions to achieve mass concentrations of 0.5%, 1%, 2%, and 4% in deionized water. This range of salinity was chosen to give changes in conductivity similar to that expected under moderate levels of hemolysis. Figure 3 shows the variation in time of the saline resistivity, $\rho = R/h$, where $R$ is the resistance in ohms and $h$ is the channel height (note that conductivity is defined as $1/\rho$). As salt is added to the reservoir, there is an initial step change in the resistivity followed by more gradual asymptotic behavior as the salt dissolves. The conductivity probe measurements were made just before each stage of salt addition, and we see that they agree well with the LCR meter readings. The measurements also match well with the known conductivity of KCl solutions (*17*).

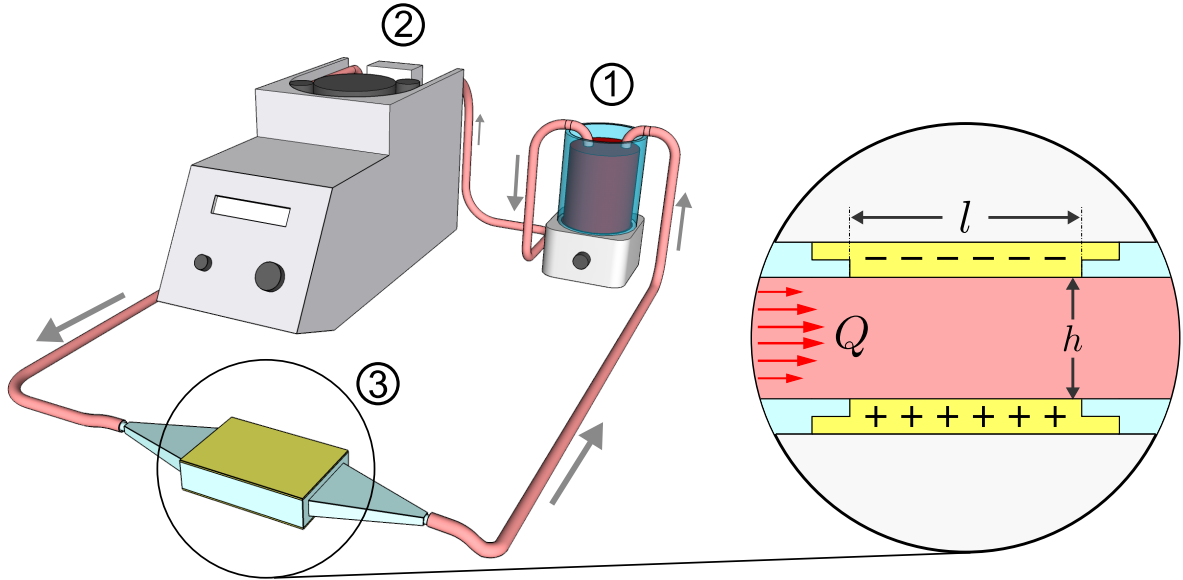

Figure 2: Experiment schematic. Flow path: (1) magnetic stirrer with open reservoir; (2) peristaltic pump; (3) 3D printed channel with a conductive floor and ceiling in the test section.

The system was then tested using porcine blood. Measuring the absolute resistivity of blood can be challenging, in that it can act as a dielectric (*15*), and also the flow shear can align the orientation of the red blood cells and make the electrical properties of blood anisotropic (*16*). Since we are only interested in measuring relative changes in blood resistivity for a given flow condition, these effects can be ignored. To control the level of damage, the blood was split into two separate 500 mL samples. One sample was left pristine while the other sample was mechanically damaged using an immersion blender (77% hemolyzed). The undamaged sample was used as the starting fluid in the flow loop, and then every 10 min 50 mL of the damaged blood sample was added into the stirring reservoir to increase the hemolysis levels in gradual steps. Samples of 1.5 mL were extracted twice at each blood damage level for direct hemolysis measurements using the spectrophotometer as described in the methods section.

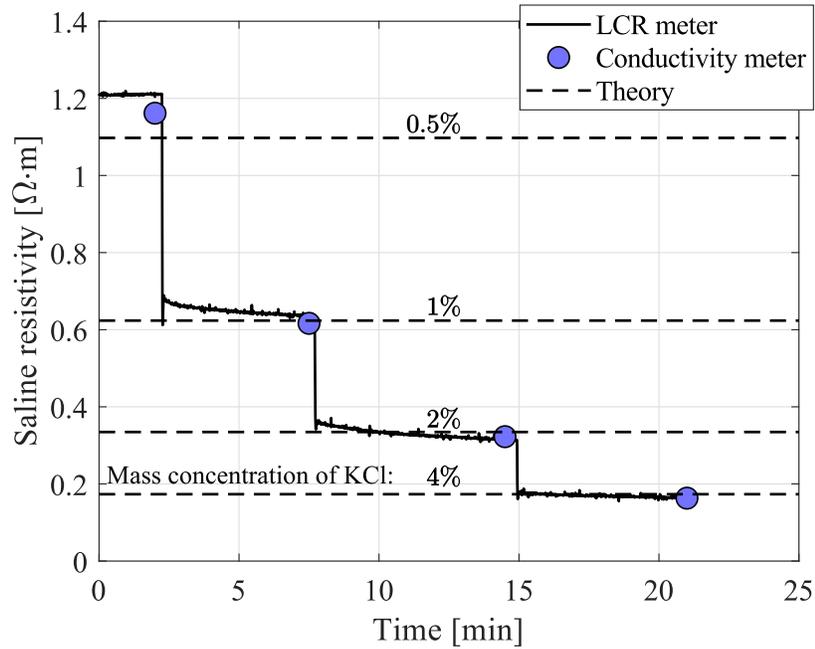

Figure 3: Saline resistivity, as measured continuously with the LCR meter and periodically with the conductivity probe, compared with the corresponding theoretical values.

Figure 4.a shows the blood resistivity measured by the LCR meter (left axis) and hemolysis percentage measured using the spectrophotometer (right axis) over the 90 minute test period. Each addition of damaged blood causes a step change in the blood resistance followed by a slower asymptotic behavior as the mixture homogenizes. Figure 4.b shows a direct correlation between the blood conductivity level ($1/\rho$) and the total hemolysis percentage.

We see that a simple conductivity cell can be used to immediately detect changes in hemolysis. The measurement is continuous, in real time, and easy to implement in clinical practice. In dialysis, for instance, the test cell can simply be incorporated in the blood flow loop and allow for immediate detection of blood damage. With further development, the test cell can obviously be made smaller and constructed using biocompatible materials, and incorporate stable, high-accuracy resistance measurement electronics.

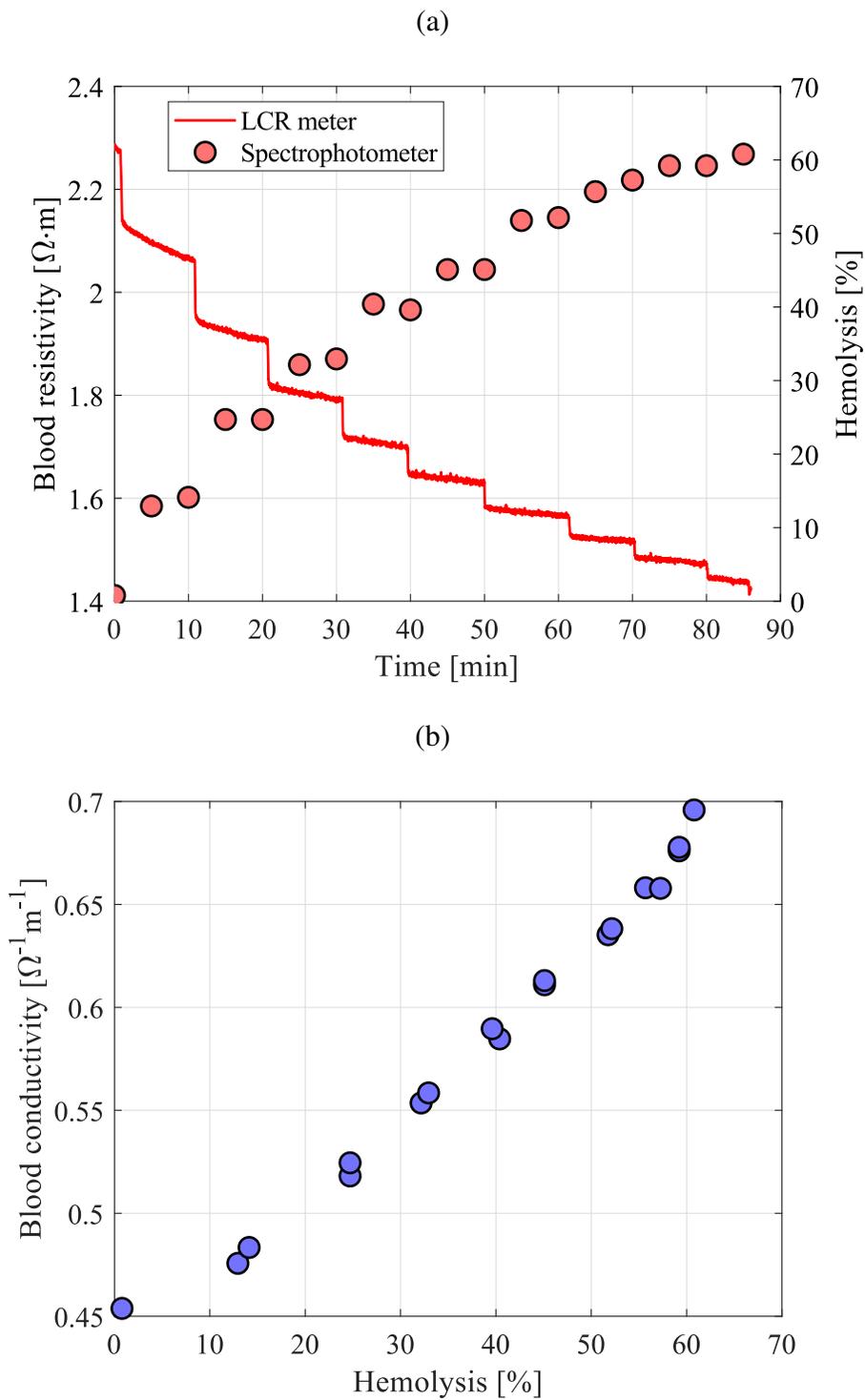

Figure 4: (a) Blood resistivity (left axis) and hemolysis percentage (right axis) as a function of time. (b) Blood conductivity as a function of hemolysis percentage.

## Materials and methods

Measurements were made in the recirculating flow facility shown in figure 2. Fluid was driven by a peristaltic pump (Cobe 043600-000) at 41.67 cm$^3$/s through a test channel and an open reservoir equipped with a magnetic stirrer (Sargent-Welch). The channel was custom designed and 3-D printed from a waterproof photopolymer (Watershed 11122XC) at the W.M. Keck Center for 3D Innovation at the University of Texas El Paso. The channel test section had length $l = 66$ mm, height $h = 7.6$ mm, and width $w = 25.4$ mm and was equipped with an electrically conductive brass floor and ceiling for flow resistance measurements. The open reservoir station allowed for direct conductivity measurements of the recirculating fluid, sample extraction, and mixing in other materials.

During the tests, the resistance of the fluid was continuously measured with an inductance-capacitance-resistance meter, or LCR meter (Keysight Technologies E4980AL) which can continuously read resistance with 0.1% accuracy. Measurements were recorded via LabView at 2 Hz, and tests were conducted for up to 90 minutes. Direct conductivity measurements were also made with a more conventional conductivity probe (Hach HQ14D), accurate to 0.5%, to validate the LCR meter readings.

Two working fluids were used for this study. First, a potassium chloride (KCl) based saline solution was used to validate our resistance measurements, which was made by mixing a known mass of KCl (EMD PX1405-1) measured via a precision scale (VWR 1002E) into room-temperature deionized water. Second, we used room-temperature porcine blood for testing hemolysis. The blood was purchased fresh through Lampire Biological Laboratories where it was obtained from healthy adult animals of unspecified gender with an added anticoagulant heparin. Hemolysis levels were determined by measuring the relative levels of free hemoglobin in the blood plasma using a spectrophotometer (Beckman Coulter DU730).

The procedure to obtain hemolysis percentage measurements of blood samples using a spectrophotometer is as follows.

[1] Prepare a Drabkin's Solution by combining Drabkin's Reagent (Sigma Aldrich D5941) to 1 L of deionized water.

[2] To lyse the red blood cells, prepare a separate solution containing 100 mL of the Drabkin's solution from Step 1 and 0.05 mL of 30% Brij 35 Solution (Sigma Aldrich B4184).

[3] Set aside two 1.5 mL samples of undamaged blood into centrifuge tubes (samples measured via Eppendorf 5 mL adjustable volume pipette). Then, acquire and similarly store blood samples during the experiment.

[4] Centrifuge the blood samples for 3 minutes at 6000 RPM to separate the red blood cells from the plasma. Leave one of the two 1.5 mL undamaged blood samples uncentrifuged.

[5] In spectrophotometer cuvettes, mix 2 mL of the Drabkin's Solution with 8 $\mu$L of plasma from the centrifuged blood samples (extracted with Eppendorf 10 $\mu$L adjustable volume pipette).

[6] Mix 8 $\mu$L of the undamaged/uncentrifuged blood sample from Step 4 with 2 mL of the Drabkin's + Brij 35 Solution from Step 2 into a spectrophotometer cuvette.

[7] Ensure that all cuvette samples are well mixed and allow to rest for 15 minutes.

[8] Zero the spectrophotometer using *only* the Drabkin's Solution (2 mL) in a cuvette, this serves as the "blank" sample.

[9] Using the spectrophotometer, measure and record the baseline reference case of the cuvette with the original undamaged blood sample from Step 5. Denote as $A_0$, where $A$ is the

absorbance at a wavelength of 540 nm.

[10] Similarly, measure and record the fully damaged reference case from the lysed blood sample made in Step 6. Denote as $A_\infty$.

[11] Lastly, measure and record the spectrophotometer readings from all of the experimental samples from Step 5.

[12] The relative hemolysis of a given sample is given by $(A - A_0)/(A_\infty - A_0)$.

# Acknowledgments

Thanks to Lena Dubitsky, Fitsum Petros, Madelyn Baron, Fernando Eugênio de Oliveira Xavier, and Gustavo de Menezes Geraldo for their work on various preliminary aspects of the project. The project was funded by the Princeton Helen Shipley Hunt Fund.